\documentclass{aa}

\usepackage{psfig,latexsym,amssymb,times}   
\usepackage[english]{babel}
\begin{document}

\title{A search for Warm-Hot Intergalactic Medium features in the X-ray spectra
of Mkn\,421 with the XMM-Newton RGS}

\author{M.~Ravasio\inst{1}
\and G.~Tagliaferri\inst{1}
\and A.M.T.~Pollock\inst{2}
\and G.~Ghisellini\inst{1}
\and F.~Tavecchio\inst{1}
}

\offprints{G.~Tagliaferri (tagliaferri@merate.mi.astro.it)}
\institute{INAF-Osservatorio Astronomico di Brera, Via Bianchi 46, I-23807 Merate, Italy 
\and European Space Astronomy Centre, Apartado 50727, 28080 Madrid, Spain }
\date{Received ....; accepted ....}
\titlerunning{A XMM-RGS search for WHIM features in the X-ray spectra of Mkn421}
\authorrunning{Ravasio et al.\ }
\abstract{We present the high-resolution X-ray spectra of Mkn\,421 obtained in November 2003
with the RGS aboard the XMM-Newton satellite. This Target of Opportunity observation was triggered 
because the source was in a high state of activity in the X-ray band. These data are compared with
three archival RGS observations of the same source performed in November and December 2002 and one
in  June 2003. We searched for the presence of absorption features due to warm-hot intergalactic 
medium (WHIM). We identify various spectral features, most of which are of instrumental origin.
With the sensitivity provided by our spectra we were able to identify
only two lines of astronomical origin, namely features at 
$23.5$ \AA, probably due to interstellar neutral oxygen absorption, and at
$21.6$ \AA, which corresponds to a zero-redshift OVII K$\alpha$ transition. For the latter,
we derive an upper limit to the gas temperature, which is consistent with WHIM,
of a few times $10^5$ K, if the gas density has a value of $n_e \gtrsim 10^{-5}$ cm$^{-3}$.

}
\maketitle
\section{Introduction}
The baryon density at $z>2$ calculated from observations of 
hydrogen and helium absorption lines 
in the Ly$\alpha$ forest (Rauch et al. 1997)
is in good agreement  with
standard Big--Bang nucleosynthesis predictions (Burles \& Tyler 1998). 
This is not true at lower redshifts: at $z \sim 0$, all current analyses
indicate that, after adding the well--observed contributions,
the local baryon density is
$\Omega_{bar,obs} \sim 0.007$ (see e.g. Fukugita, Hogan \& Peebles 1997),
much lower than the predicted value 
$\Omega_{bar,n-syn} \sim 0.035-0.04$ (Burles \& Tytler 1998).\\
High--resolution, large--scale  hydrodynamic simulations of galaxy formation
have been used to predict the baryon distribution
at the present epoch and at moderate redshift.
 The main result of such simulations is
that approximately  30\%--40\% of the baryons in the present--day universe 
should reside in a Warm--Hot Intergalactic Medium (WHIM), shock--heated to
temperatures of $10^5 - 10^7$ K. 
Most of these warm--hot baryons seem to reside in diffuse filamentary 
large--scale structures with overdensities of 10--30 
and not in virialized objects such as galaxy groups (Cen \& Ostriker 1999;
Dav\'e et al. 2001).\\
Numerical simulations have demonstrated that H--like and He--like
ions of the heavy elements composing the WHIM can give rise to absorption lines 
in the  soft X--ray spectra of background sources.
OVII, OVIII and NeIX should dominate the relative abundance distributions
of collisionally ionized and photoionized+collisionally ionized gas 
over a broad range of temperatures
($5\times 10^5$--$10^7$ K, Nicastro et al. 1999),
where the WHIM distribution peaks (Dav\'e et al. 2001;
Fang \& Canizares 2000). 
Absorption lines at 21.6 \AA, 18.97 \AA\, and 13.4 \AA, respectively,
should then be observable in the soft X--ray spectra 
of strong background sources
(see e.g. Aldcroft et al. 1994; Hellsten, Gnedin \& Miralda--Escud\'e 1998;
Perna \& Loeb 1998; Fang \& Canizares 2000).\\
Absorption lines such as the doublet  
of OVI (1032 \AA\, and 1038 \AA) should  be detectable also in 
the optical--UV band
(Mulchaey et al. 1996; Cen et al. 2001), for temperatures
of $1-5\times 10^5$ K, tracing the low--temperature tail of the
WHIM distribution. Current far--UV observations have proved the existence
of such a low--temperature component with the detection
of OVI absorption lines up to $z\sim0.2$ 
(e.g. Sembach et al. 2000; Tripp et al. 2001).\\
Soft X--ray spectra provide an even better opportunity than the UV spectra:
the detection and the study of these components 
is needed for the proper understanding of large and small--scale structures
in the Universe, providing independent constraints 
on cosmological parameters.\\
The predicted highly ionized gas, however, has been poorly studied 
so far, because of instrumental limitations.
New spectrometers aboard Chandra (HRCS/LETG) 
and XMM--Newton (RGS) have increased the sensitivity and the resolution of the 
X--ray observatories slightly beyond the WHIM detection limit.
At present, only  the  strongest of these systems (EW\,$>10$ m\AA)          
have been detected against the spectra of very bright background sources
(Nicastro et al. 2002; Mathur, Weimberg \& Chen 2002; Fang et al. 2002;
Fang, Sembach \& Canizares 2003; Cagnoni et al. 2004; Nicastro 2003).
Furthermore, only two of them were identified as signatures  of the WHIM 
 outside the Local Group: an absorbing system at $z\sim0.05$
 toward 3C\,273 (Fang et al. 2002) and  one at $z \sim 0.01$ 
toward Mkn 421 (Cagnoni et al. 2002; Nicastro et al. 2003).\\
High--resolution observations of Mkn 421
have already been performed with Chandra (Nicastro et al. 2001;
 Nicastro 2003) 
and with XMM--Newton  (Cagnoni 2001; de Vries et al. 2003).
The [18--24] \AA\, spectrum of Mkn 421 changed during the 
two Chandra observations of 2000.
In one, the source was very bright and
no absorption or emission lines were detected. In the other, 
negative and positive residuals from the best--fit power--law model 
were observed. Nicastro et al. (2001) tentatively identified two different
absorbing/emitting systems and proposed that they are intrinsic 
to the nuclear environments, becoming fully ionized - and thus transparent -
as the source brightens.\\
Further Chandra observations, performed while the source
 was in a very bright phase, allowed Nicastro (2003) to claim
the presence of three absorbing systems located at $z\sim0$, $z\sim 0.011$
and at $z\sim0.03$, respectively. 
The first was identified as the WHIM inside the Local Group,
while the last was explained as Mkn\,421 intrinsic absorption. 
The system st $z\sim0.011$, was interpreted as the WHIM outside 
the Local Group.\\
Even the RGS spectrum of  Cagnoni (2001) showed 
two absorbing systems, one inside the Local Group 
and one at $z\sim0.01$, contributing an OVII K$\alpha$ absorption line
at $\sim 21.8$ \AA. 
de Vries et al. (2003), instead, concentrated on the [22--24] \AA\, 
range of the  RGS data and found evidence for a 23.5 \AA\,
interstellar neutral Oxygen (1s--2p) absorption feature.
 Furthermore, they showed a feature at 22.77 \AA, 
which they argued to be a non--Galactic OVI blend.\\
In this paper we present the results of a XMM--Newton RGS  observation
of Mkn 421 performed as part of a Target of Opportunity (ToO) program to
observe Blazars in high state of activity (see Tagliaferri et al. 2001).
We triggered the pointing because 
the All Sky Monitor aboard {\it Rossi} XTE had been reporting 
high X--ray fluxes from Mkn 421 for several days.
We compared our observations with 3 archival RGS spectra of Mkn 421 
taken in November and Dicember 2002 and with one taken in June 2003, as part
of two different calibration campaigns.\\
\begin{table*}[t]
\begin{center}
\begin{tabular}{cccccc}
\hline
\multicolumn{6}{c}{\bf XMM-Newton RGS}\\
\hline
Revolution & Obs. Id. & \multicolumn{2}{c}{Start time} & Total exposure$^a$ & Net exposure  \\
           &          &  Day & Hour   &  ($10^4$ s)    & ($10^4$ s)   \\
\hline
 0532      & 0136540301 & 04/11/2002 & 00:44:59&  2.3        &  2.1 \\
\hline
 0532      & 0136540401 & 04/11/2002 & 07:41:43 & 2.3        &  2.1 \\
\hline
 0546      & 0136541001 & 01/12/2002 & 22:59:25 & 7.1        &  5.8 \\
\hline
 0637      & 0158970101 & 01/06/2003 & 12:48:50 & 4.3        &  3.0 \\ 
\hline
 0720      & 0150498701 & 14/11/2003 & 16:14:04 & 5.7        &  4.7 \\
\hline
\end{tabular}
\caption{Log of the XMM--Newton RGS observations of Mkn 421 taken in 2002 and 2003
that were used during this analysis.
A further observation of Mkn421 0537\_0136540701 was also performed during the
cooling of RGS1 which we have not used.
$^a$: total amount of Good Time Intervals.
$^b$: the reprocessing of these data failed; we exclude them from 
further analysis.}
\label{tab1}
\end{center}
\end{table*}

In the following sections, we briefly describe the observations
and show the results of the spectral analysis 
performed on the whole RGS energy range. 
Then we concentrate on
smaller energy ranges, looking carefully for the presence
of absorption features whose reality is checked in two ways:
first, by comparing the
Mkn 421 spectra with the smooth X-ray spectrum of the Crab nebula
and second, by paying particular attention to the raw data
to check for instrumental effects.
Occasional hot or cool pixels are routine in CCD data as a result of
cosmic--ray damage and other effects and are treated as part of
normal data analysis procedures.
Such defects are usually confined to single pixels and are thus
significantly narrower than the instrumental line response
that is principally caused by scattering from the gratings.
Finally, we discuss the results proposing an identification
for the possible non--Galactic absorption lines.

\section{The XMM--Newton observations and data reduction}

The XMM--Newton X--ray  payload consists of three Wolter type--1 telescopes, 
equipped with 3 CCD cameras for X--ray imaging,
moderate resolution spectroscopy
and photometry (EPIC). Two of these telescopes (those carrying the 
MOS cameras) are also provided  with high resolution 
Reflection Grating Spectrometers  (RGS--1 \& RGS--2), 
operating in the range [0.33--2.5] keV (5--38 \AA). 
Each RGS unit deflects half of its telescope beam, dispersing
the striking X--ray light at a wavelength--dependent angle, thus
providing a spectral resolution of $E/\Delta E \sim 100 - 500$ (FWHM).
After the launch, however,  failures in the read--out electronics
of the CCD--7 (RGS--1) and CCD--4 (RGS--2),
covering the [10.5--14] \AA\,  and the [20.1--23.9] \AA\, ranges, respectively,
reduced by a factor of 2 the RGS effective area at these wavelengths.\\
Mkn 421 was the target of a RGS calibration campaign in November and December
2002, aimed at improving the instrumental performances by lowering
the operating temperature. The benefits of the  cooling manifested 
as a dramatic reduction of hot columns and flickering pixels,
as well as an increase of the Charge Transfer Efficiency
(see the movies at the XMM--Newton site).
The RGS--1 and RGS--2  were cooled in the night between
November 13--14 and 3--4,
respectively. A third observation of Mkn 421 was carried out
on December 1, 2002.\\
A further calibration campaign was performed in June 2003. The observational
time was fractioned in several short pointings: in our analysis 
we included only the longest one (lasting $\sim 43$ ks).\\
Finally, we triggered a 50 ks ToO observation on  12 November 2003,
but, because of the intense Solar activity, we had to 
re--schedule it two days later, when, according to the ASM, 
the source was still in a high state.
The log of the analyzed observations
is given in Table \ref{tab1}.
We excluded from the  analysis the data collected 
during the RGS--1 cooling night (14/11/2002),
because of reprocessing failures.\\
We reprocessed  the data  using the XMM--Newton Science Analysis System (SAS)
5.4.1 and the same calibration files used
by the XMM--Newton Survey Science Centre (SSC) 
in the  standard Pipeline Processing (PPS files).\\
Since the XMM--Newton instruments are affected by periods 
of high background activity induced by solar flares, 
we extracted the light curves of both instruments
from a background region of the CCD--9, which is the closest to the instrument
axis and the most susceptible to proton events (Snowden et al. 2002).
We then excluded the flaring time intervals
(net exposures are reported in Table \ref{tab1}).
After this filtering, we re--extracted the source and
background spectra within the 95\% and outside the 98\% of the PSF, 
respectively. Since the RGS wavelength calibration 
is strongly position--dependent,  we fixed the source position 
to the VLBI coordinates (Ma et al. 1998). Because of significantly
better statistics, we focused our analysis on the first--order data only.

\section{Looking for lines in various segments of the spectra}

Below 0.5 keV there are still some uncertainties in the EPIC calibration. Thus,
before looking for the WHIM signature, we examined the full range of the RGS spectra 
to derive and compare the Mkn 421 spectral shape, with the values
obtained with EPIC (see Ravasio et al. 2004). Furthermore, we checked the 
cross--calibration between the RGS and the EPIC--PN detectors in the common energy range 
([0.6--1.77] keV), where both instruments should be properly calibrated.
These results are reported in the Appendix, where in Table \ref{rgs-bband-spec} we also
give the source fluxes.

In this Section we shall look for the possible presence 
of faint absorption features in the RGS spectra,
which could be the signature of the WHIM  toward the source. 
According to the simulations on the WHIM chemical composition,
the strongest absorption lines should be the  OVII K$\alpha$ 
(21.602 \AA\, in the observer frame) and the OVIII K$\alpha$ (18.97 \AA)
(Hellsten, Gnedin \& Miralda--Escud\'e 1998).
Therefore we concentrated our analysis on small  energy ranges
(2--3 \AA\, wide) centered on these wavelengths
as well as on the [22.5--24] \AA\, range,
where the interstellar OI 1s--2p absorption line ($\sim 23.5$ \AA)
was already observed by XMM--Newton and Chandra 
toward Mkn 421 (de Vries et al. 2003)
and toward other sources (e.g. PKS 2155-304, Cagnoni et al. 2004).\\
Because of failures in the read--out electronics of the CCD--4 of the RGS--2,
covering the [20.1--23.9] \AA\, band, we focused
on the RGS--1 data, using the RGS--2 spectra, where available, to check
the reality of the possible features. We  also compared the RGS--1 spectra 
with a RGS--1 spectrum of a powerful Galactic source, the Crab.\\
Using XSPEC 11.2.0, we extracted the unbinned, unfolded spectra in the
[18--20] \AA, [21--22.3] \AA\, and [22.5--24] \AA\,
intervals and analyzed them with Sherpa 2.2.1 which  can better work
in the wavelength space. Each feature that we studied, 10 in total, is 
identified by a unique number throught the paper and in the figures.

\subsubsection{Wavelength calibration}

\begin{figure*}
\hspace{-1cm}
\hbox to \textwidth
{
\vbox{\psfig{figure=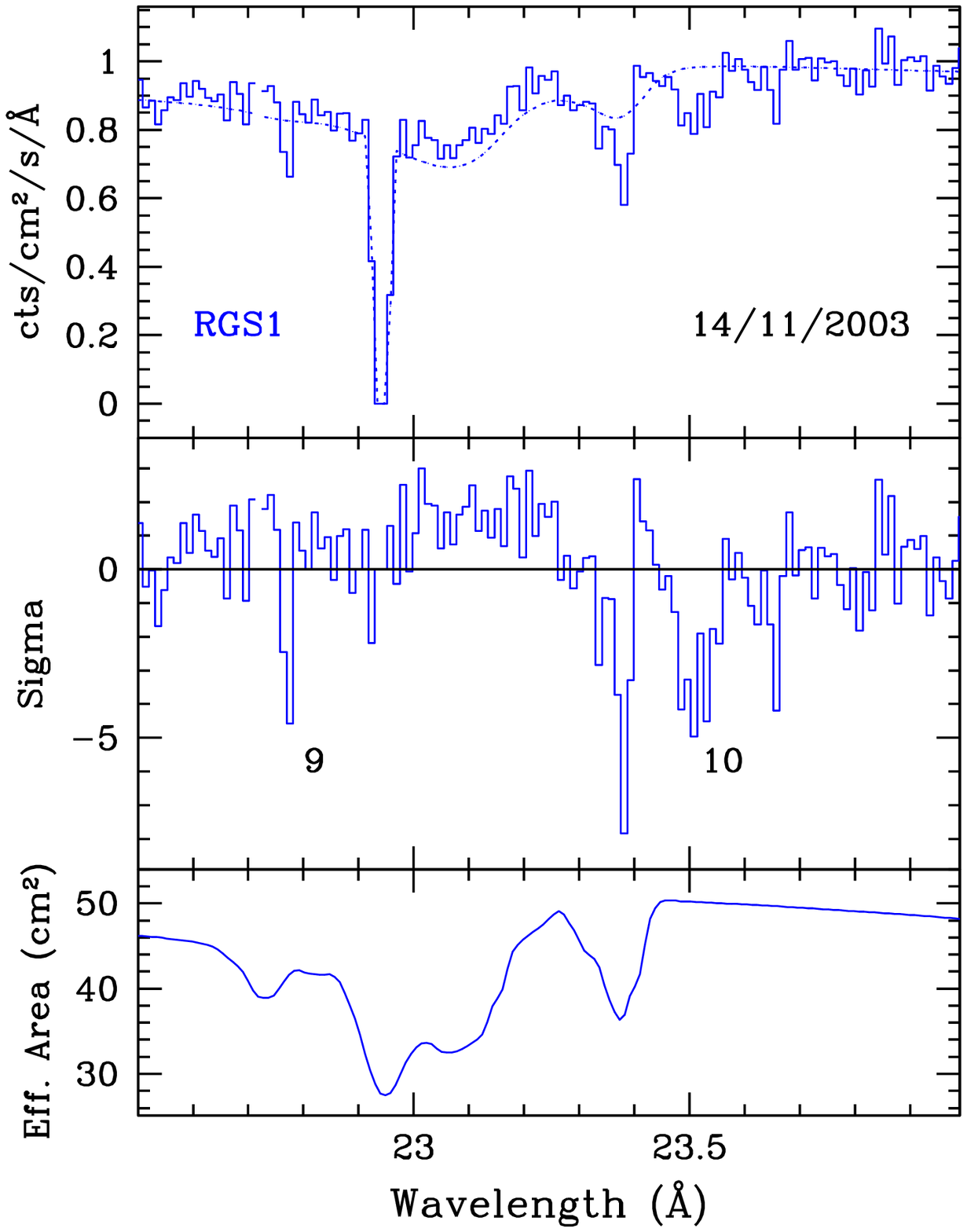,width=9.5cm}}
\hskip -0.1cm
\vbox{\psfig{figure=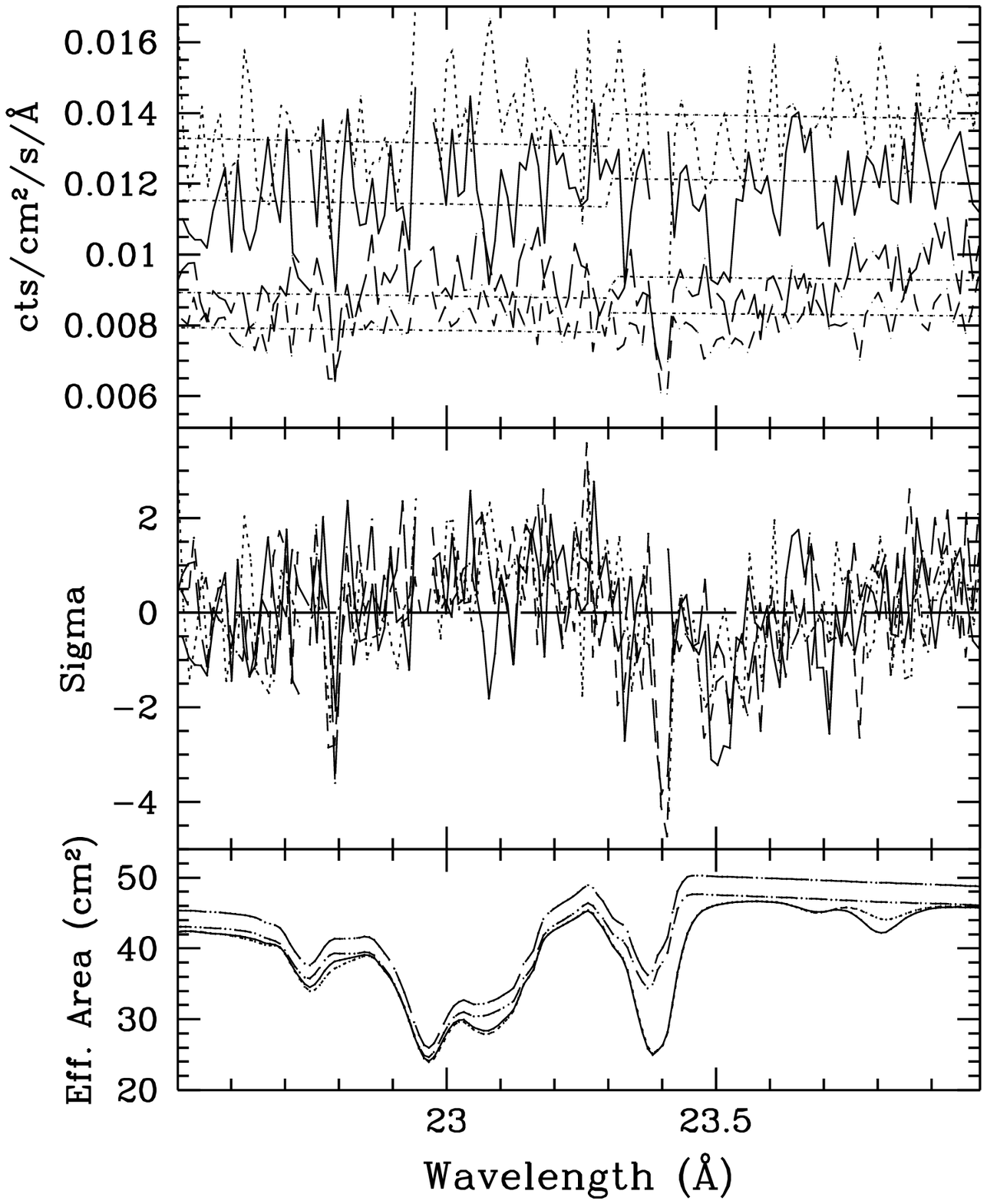,width=9.5cm}}
\hfill
}
\caption{The ToO (left) and the archival (right)
RGS--1 spectra of Mkn 421 in the [22.5--24] \AA\, band,
together with the best--fit power--law models (upper panels).
In the right panel, 
the solid  and the dotted lines refer to the first and the second exposures
of November 4, 2002, respectively, the short--dashed one
represents the December 1$^{\rm st}$, 2002 data while the long--dashed
data are those taken in June 2003.
In the middle panel we report the residuals left by the best--fit models
and in the bottom panel the RGS--1 effective areas.}
\label{2224-calib}
\end{figure*}

We used the interstellar neutral oxygen feature  at $\sim23.5$ \AA,
(OI 1s--2p, $1s^2 2s^2 2p^4~^3P^0 - 1s 2s^2 2p^5~^3P^0$) 
to determine the absolute line position in the RGS--1 spectra,
which is fundamental to obtain reliable redshifts of the 
possible absorbing systems.\\
This line was already observed 
toward Mkn 421 by XMM--Newton and by Chandra (de Vries et al. 2003)
as well as toward other sources, such as PKS\,2155-304 (Nicastro et al. 2002;
Cagnoni et al. 2004) or H 1821+643 (Mathur et al. 2003).\\
In  Fig. \ref{2224-calib} we show the [22.5--24] \AA\, 
RGS--1 spectra of our ToO observation (left) and
of the archival data  (right).
We plot the spectra and the best--fit power--law models
(upper panel), the residuals (mid panel) and the RGS--1
effective areas during each  XMM--Newton observation (bottom panel).
For the archival data, we plot
the first and the second exposures of November 4, 2002
as solid and dotted lines, respectively,
 the December 1$^{\rm st}$, 2002 data as a short--dashed line
and the June 1$^{\rm st}$ data as a long--dashed line.
Residuals at $\sim 23.5$ \AA\, can be observed in the middle panels
of Fig. \ref{2224-calib}, which are very likely produced by the
interstellar neutral oxygen. Firstly, we fitted the ToO observation
with a power--law + one Gaussian, then we simultaneously fitted the four
archival spectra. Finally we fitted the five spectra together.
The best--fit line position during the ToO observation is 
$23.510\pm0.007$ \AA, which is consistent with the archival data
result ($23.507\pm0.013$ \AA). Fitting simultaneously all the spectra
we obtained $\lambda = 23.510\pm0.007$ \AA.\\
This value is slightly higher than the theoretical position found by 
Mc Laughlin \& Kirby (1998; $\lambda=23.467$ \AA), but it is 
consistent with the results of many authors, obtained through
different experimental techniques (see Table \ref{calib}).
\begin{figure}
\hspace{-1cm}
\psfig{figure=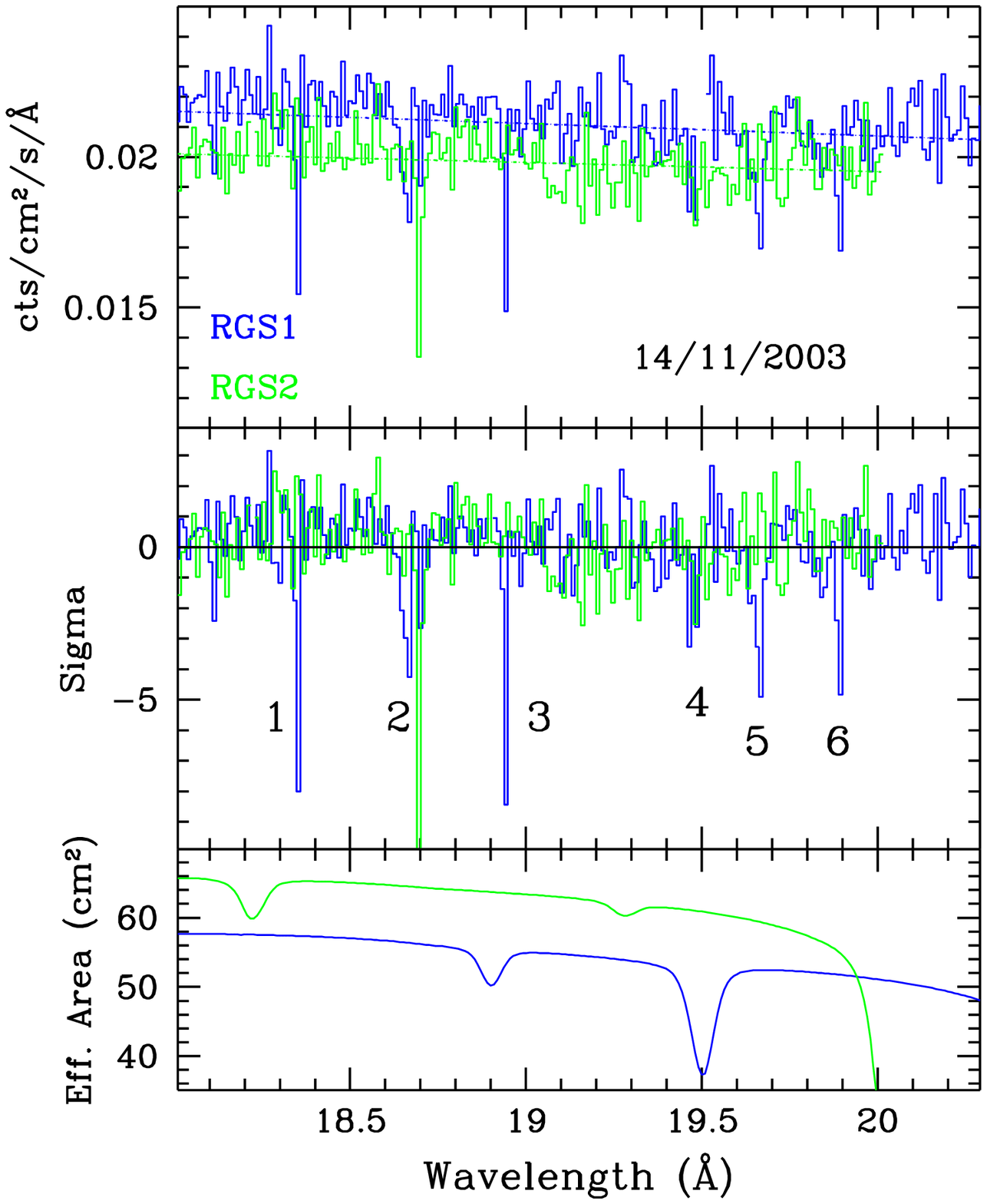,width=9.5cm}
\caption{The RGS--1 (dark gray data) and the RGS--2
[18--20.3] \AA\, spectra (light gray data) of Mkn 421 
during the ToO observation of November 2003.
We also show the respective best--fit power--law models (dashed lines).
In the mid panel we report the residuals left by the best--fit models
and in the bottom panel the effective areas. 
The RGS--2 effective area decays above $\sim 20$ \AA\, because of the 
CCD--4 failure.}
\label{1820-rgs-too}
\end{figure}          
\begin{table}
\begin{center}
\begin{tabular}{ccc}
\hline
$\lambda$ (\AA)& Method &  Ref.\\
\hline
23.467  & Theoretical & 1  \\
$23.489\pm0.004$  & Experimental & 2 \\
$23.536\pm0.002$  & Experimental & 3 \\
$23.52\pm0.03$  & Chandra, X0614+091 & 4 \\
$23.509^{+0.008}_{-0.018}$ & Chandra, PKS 2155-304 & 5 \\
$23.50 \pm 0.01$  & XMM & 6 \\
$23.510\pm0.015$ & XMM, PKS 2155-304 & 7 \\
$23.510\pm0.007$ & XMM, Mkn 421 & ToO obs.\\
$23.507\pm0.013$ & XMM, Mkn 421 & Arch. data \\
$23.510\pm0.013$ & XMM, Mkn 421 & Total \\
\hline
\end{tabular}
\caption{Neutral oxygen wavelengths (1s--2p transition).
1) Mc Laughlin \& Kirby, 1998;
2) Krause 1994;
3) Stolte et al. 1997;
4) Paerels et al. 2001;
5) Nicastro et al. 2002;
6) de Vries et al. 2003;
7) Cagnoni et al. 2004.}
\label{calib}
\end{center}
\end{table}
In particular, it is in good agreement ($< 1\, \sigma$) 
with other XMM--Newton and Chandra observations. Therefore,
we shall not apply wavelength corrections to our data.
\subsection{The [18--20.3] \AA\, spectra}
\begin{figure*}
\hspace{-1cm}
\hbox to \textwidth
{
\vbox{\psfig{figure=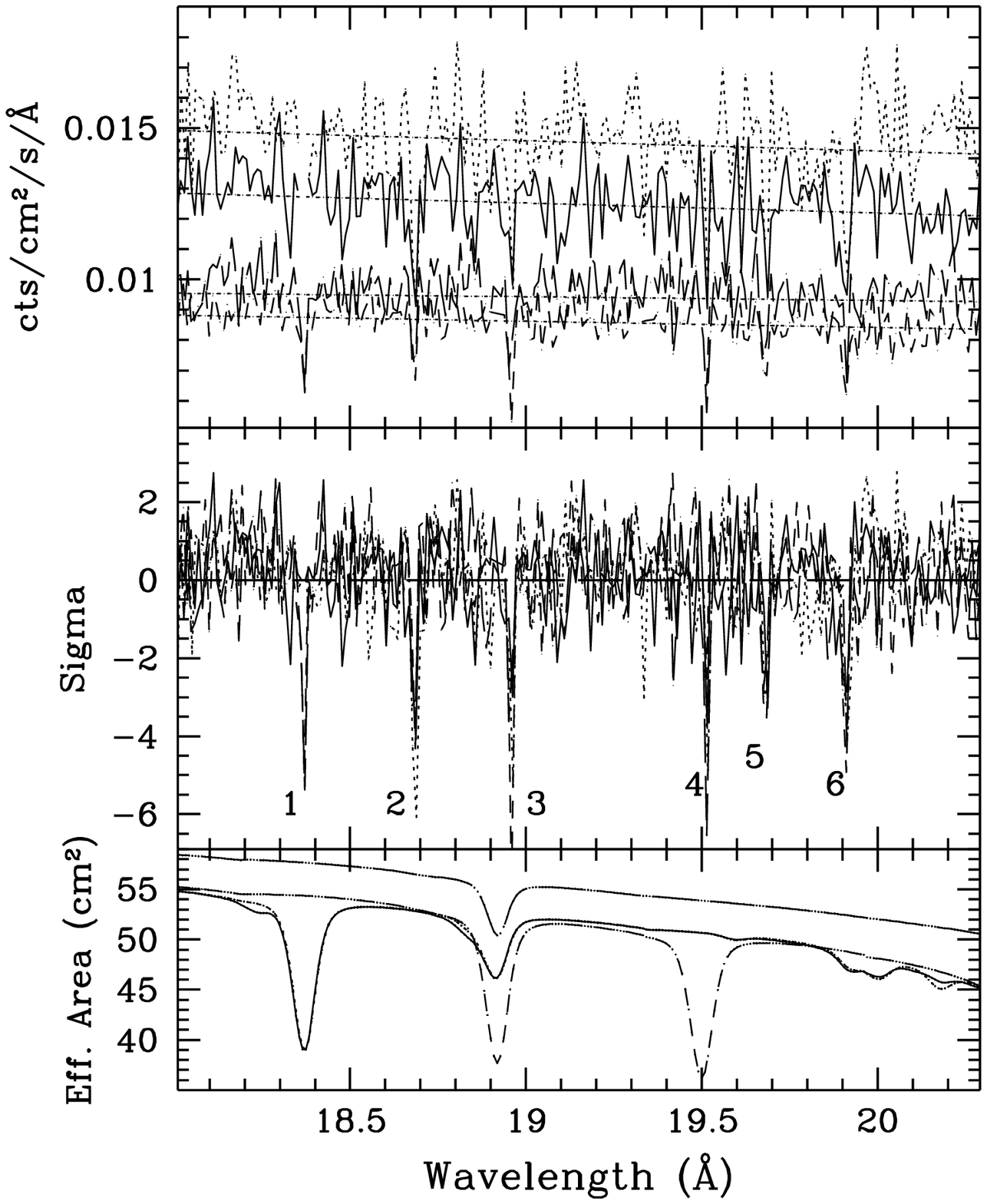,width=9.5cm}}
\hskip -0.1cm
\vbox{\psfig{figure=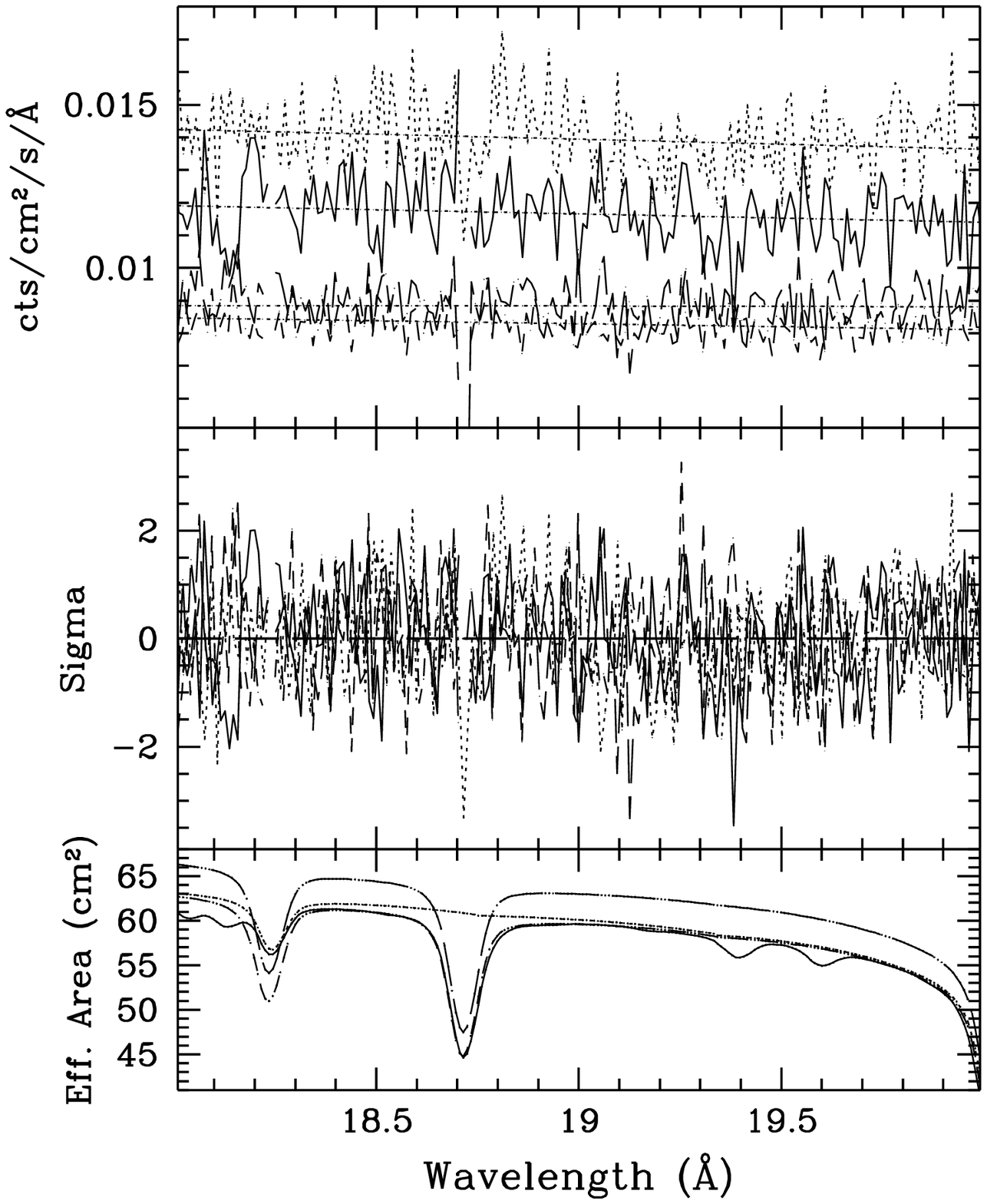,width=9.5cm}}
\hfill
}
\caption{The RGS--1 [18--20.3] \AA\, (left)
and the RGS--2 [18--20] \AA\, (right) archival spectra of Mkn 421,
together with the best--fit power--law models (upper panels).
The solid and dotted data refer to the first and the second exposures
of November 4, 2002, respectively, the short--dashed ones
represent the December 1$^{\rm st}$, 2002 data while the long--dashed
ones the June 2003 data.
In the mid panel we report the residuals left by the best--fit models
and in the bottom panel the effective areas. We limit the RGS--2
plot to the [18--20] \AA\, range because of the CCD--4 failure.}
\label{1820-rgs-arch}
\end{figure*}

\begin{figure}
\begin{center}
\psfig{figure=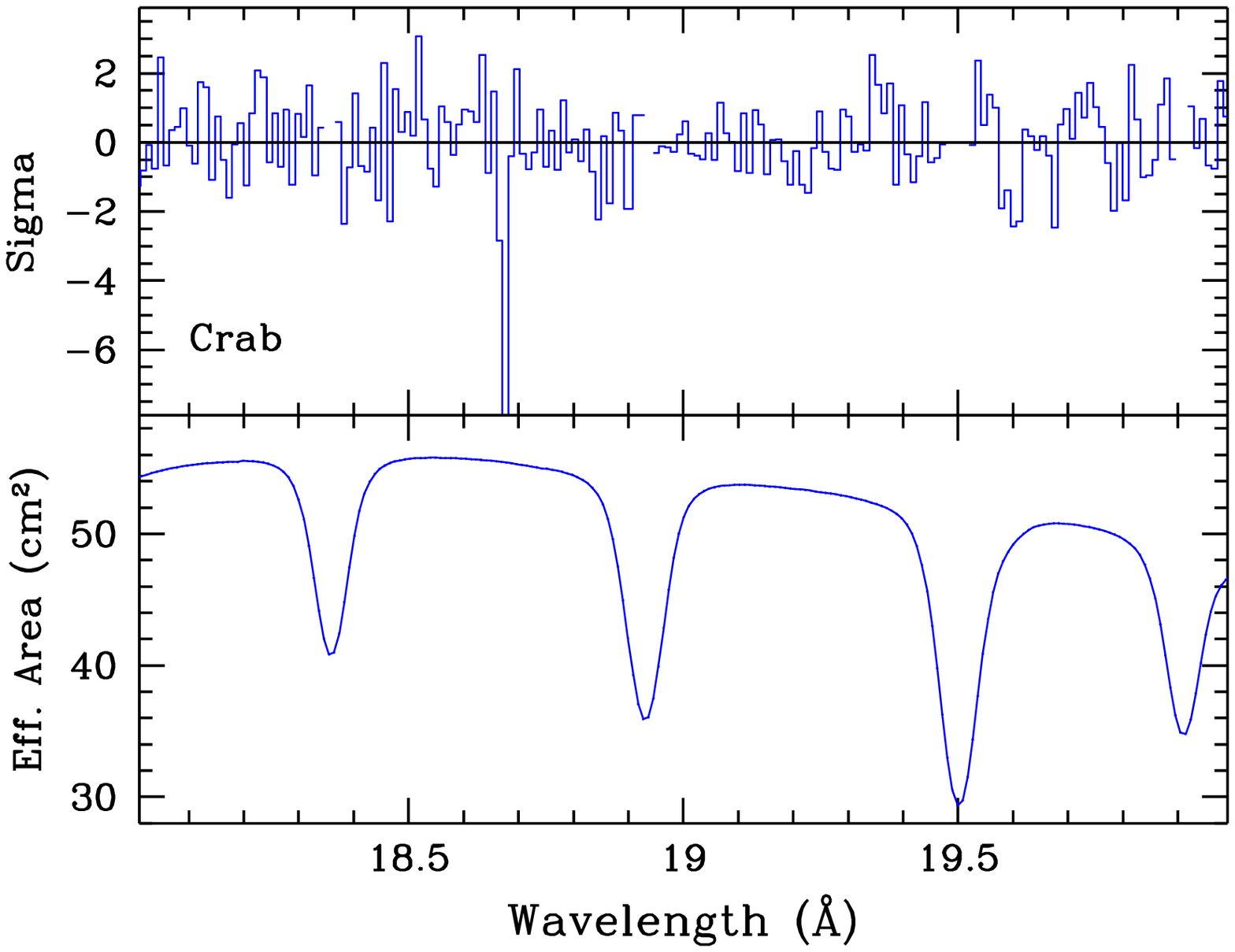,width=8.5cm}
\caption{Residuals left by the best--fit power--law model 
to the [18--20] \AA\, RGS--1 spectrum of the Crab (upper panel).
In the lower panel we show the relative RGS--1 effective area.}
\label{1820-crab} 
\end{center}
\end{figure}

In this energy range we can directly compare 
the RGS--1 and  RGS--2 data.
In Fig. \ref{1820-rgs-too} we show the RGS--1 (dark gray lines)
and the RGS--2 spectra (light gray lines) of the ToO 
observation, together with the best--fit absorbed power--law model 
(dotted lines). We report also 
the residuals (in terms of sigmas, mid panels)
and the effective area of the instruments (bottom panels) during the exposure.
In Fig. \ref{1820-rgs-arch}, we show the same
plots for the archival observations. We display as solid and dotted lines
the first and the second exposures of November 4,
2002, respectively, as a short--dashed line the December 1$^{\rm st}$, 
2002 data and as a long--dashed line the June 1$^{\rm st}$ data.\\
Several features  with significance $\gtrsim3 \, \sigma$ are shown in all
the RGS--1 spectra. Two of them, the numbers 1 and 4, are located in some 
of the archival observations
at the same energies as large structures in the effective area curves.
Furthermore, they cannot be observed in the corresponding RGS--2 residuals,
in regions where the effective areas are smooth. Their origin is very likely 
instrumental and we shall exclude them  from further analysis.\\
We then fitted the RGS--1 spectra again, with  an 
absorbed power--law model plus four Gaussian profiles
to reproduce the residuals  2, 3, 5 and 6.
In Table \ref{1820-tab} we report the best--fit parameters
of each Gaussian.\\
\begin{table}[!b!]
\begin{center}
\begin{tabular}{cccc}
\multicolumn{4}{c}{RGS--1 [18--20] \AA}\\
\hline
Feature & $\lambda$ & FWHM  & k  \\
number  &  (\AA)    & (m\AA) & ( cts cm$^{-2}$ s$^{-1}$ \AA$^{-1}$)\\
\hline
\multicolumn{4}{c}{ToO observation}\\
\hline
2  & $18.661\pm0.004$ & $24^{+9}_{-6}$ & $-0.0040\pm0.0011$ \\
3  & $18.937\pm0.001$ & $8\pm1$ & $0.0111\pm0.0023$ \\
5  & $19.662\pm0.005$ & $30\pm10$ & $-0.0040\pm0.0010$ \\
6  & $19.890\pm0.001$ & $8.5^{+1.5}_{-1}$ & $-0.0098\pm0.0031$ \\
\hline
\multicolumn{4}{c}{Archival data}\\
\hline
2    & $18.683\pm0.001$ & $15.2^{+3}_{-2}$ & $-0.319\pm0.048$ \\
3    & $18.9585\pm0.001$ & $14.5\pm2$ & $-0.349\pm0.051$ \\
5    & $19.681\pm0.002$ & $15\pm3$ & $-0.266\pm0.053$ \\
6    & $19.910\pm0.002$ & $16\pm3$ & $-0.303\pm0.050$\\
\hline 
\multicolumn{4}{c}{Total}\\
\hline
2 & $18.680\pm0.004$ & $33^{+9}_{-7}$ & $-0.147\pm0.026$ \\
3 & $18.953\pm0.003$ & $28^{+5}_{-4}$ & $-0.177\pm0.029$ \\
5 & $19.676\pm0.003$ & $23^{+5}_{-4}$ & $-0.166\pm0.032$ \\
6 & $19.904\pm0.004$ & $33^{+8}_{-7}$ & $-0.128\pm0.027$ \\
\hline
\end{tabular}
\caption{Best--fit parameters of the power--law plus four Gaussians
model reproducing the [18--20] \AA\, RGS--1 spectra of Mkn 421.
The error bars refer to the 90\% confidence intervals.}
\label{1820-tab}
\end{center}
\end{table}
Besides investigating the corresponding RGS--2 spectra, we also checked
the reality of these features by studying a RGS--1 spectrum
of the Crab taken in August 8, 2002 (Obs. Id. 0153750501).
The Crab nebula, a Galactic source, has a very intense featureless power--law 
continuum and is therefore very useful to 
discriminate between the possible origins of the observed features.
We reduced the Crab data as described in the previous Sections.
In Fig. \ref{1820-crab} we report the residuals (upper panel)
given by the best--fit power--law model
in the [18--20] \AA\, range and  the relative  effective area (lower panel).\\
We summarise our results as follows:
\begin{itemize}
\item Feature 2 (18.680 \AA): we found large residuals ($5-6 \, \sigma$)
close to the observer--frame wavelength of the OVII K$\beta$ transition, 
in a smooth region of the RGS--1 effective area.
This identification, however, is still controversial.
The RGS--2 data do not unambiguously confirm the RGS--1 results.
During the first ToO observation and during the second one of November 4, 2002, 
the RGS--2 effective area was smooth 
and we observed a very narrow feature at a slightly higher wavelength 
($18.697\pm 0.002$ \AA, FWHM =$14\pm3$ m\AA\, and
$\lambda = 18.718\pm0.004$ \AA, FWHM = $17 \pm 8$ m\AA, respectively).
However, a large structure is present in the RGS--2 effective 
areas of the other observations.
Furthermore, we also observed large residuals ($\sim 8 \, \sigma$)
in the RGS--1 Crab spectrum ($18.674^{+0.002}_{-0.003}$ \AA).
This implies a Galactic or an instrumental origin:
the astronomical nature of this feature is still questionable.
\item Feature 3 (18.953 \AA): large residuals ($5-6 \, \sigma$) are observed
at the zero--redshift wavelength of the OVIII K$\alpha$ line.
This identification also is quite doubtful. This line
is located close to a RGS--1 effective area feature (at $\sim 18.9$ \AA)
and  cannot be observed in the RGS--2 spectra,
 where the effective areas are smooth.
If the RGS--2 response is well calibrated, this feature must be instrumental
and the RGS--1 residuals are probably caused by calibration uncertainties 
of the effective area structure at $\sim 18.9$ \AA.
\item Feature 5 (19.676 \AA):  this  $\sim 3 \, \sigma$ feature is observed
in all the RGS--1 spectra, but it is not present in the corresponding 
RGS--2 data. Also the RGS--1  Crab spectrum displays small residuals
($\sim 2 \, \sigma$) at $19.673$ \AA. The absence of this line in the RGS--2 
suggests a very likely instrumental origin.
\item Feature 6 (19.904 \AA): the RGS--1 residuals
($\sim 4 \, \sigma$) are not observed in the RGS--2 spectra. 
In this case, the comparison with the Crab spectrum is useless, 
since the corresponding effective area is characterized by a large feature. 
This line is probably instrumental.
\end{itemize}
\subsection{The [21--22] \AA\, spectra}
\begin{figure*}
\hspace{-1cm}
\hbox to \textwidth
{
\vbox {\psfig{figure=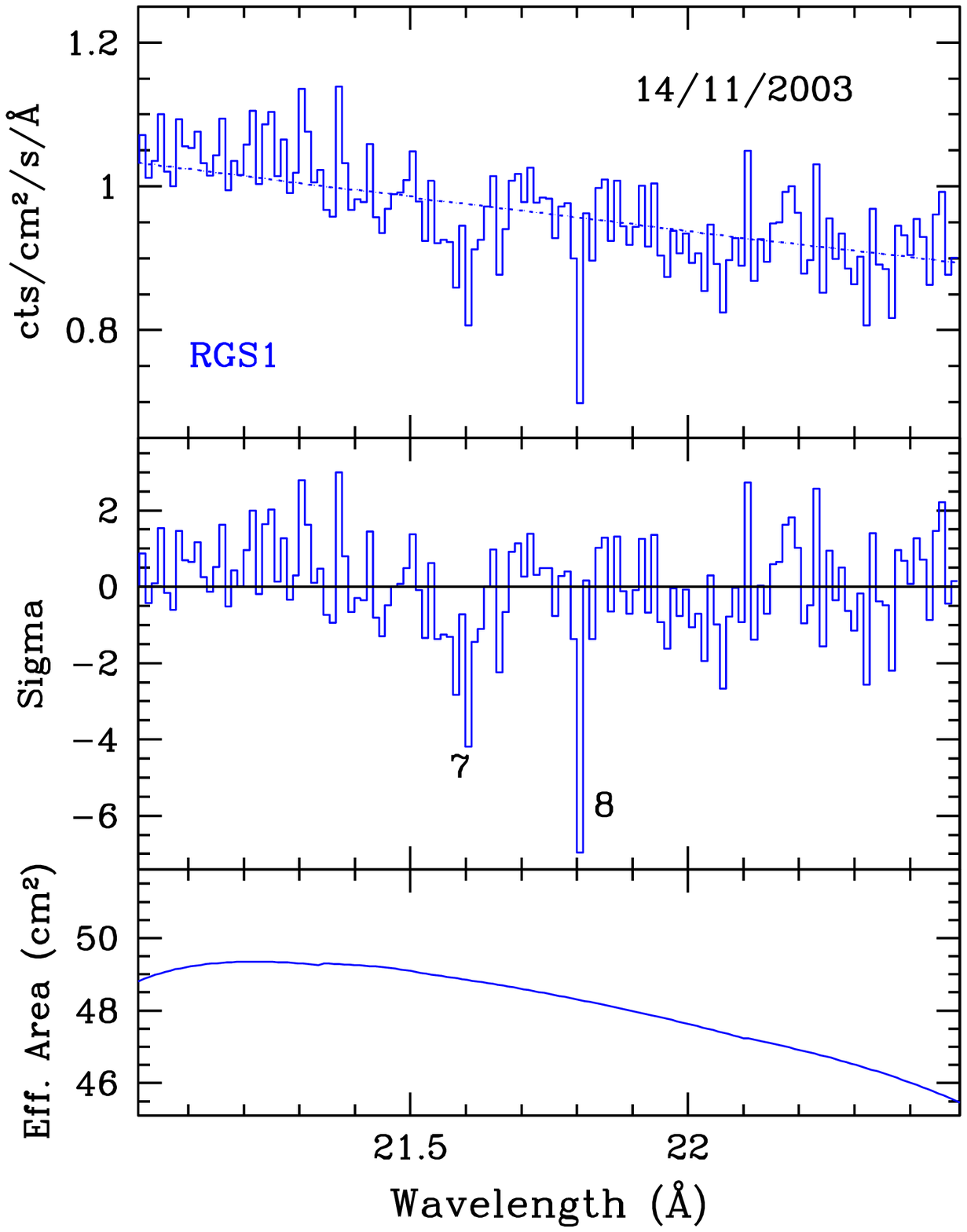,width=9.5cm}}
\hskip -0.1cm
\vbox{\psfig{figure=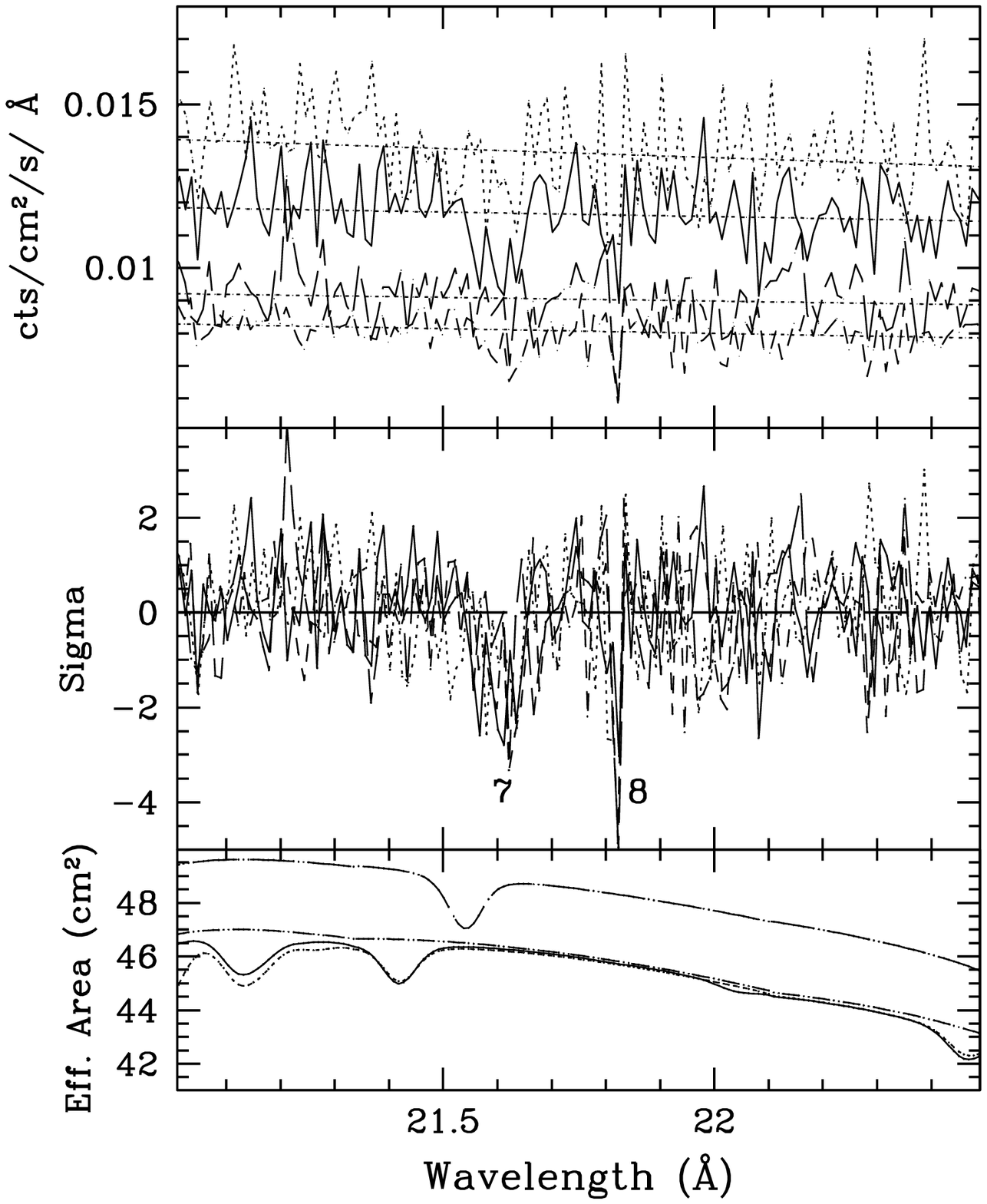,width=9.5cm}}
\hfill
}
\caption{The RGS--1 spectra of Mkn 421 in the [21--22] \AA\, range.
In the left panel we show the November 2003 ToO spectrum and in the 
right panel the archival data (upper panels), together with the
best--fit models. On the right,
we show with a solid and a dotted line the two exposures
of November 4, 2002,
with a short--dashed line the December 1$^{\rm st}$, 2002 data and
with a long--dashed line the June 2003 data.
In the middle panels we display the residuals left 
by the best--fit models and in  the lower panels 
the RGS--1 effective areas of the RGS--1.}
\label{2122-spec}
\end{figure*}

\begin{figure}
\begin{center}
\psfig{figure=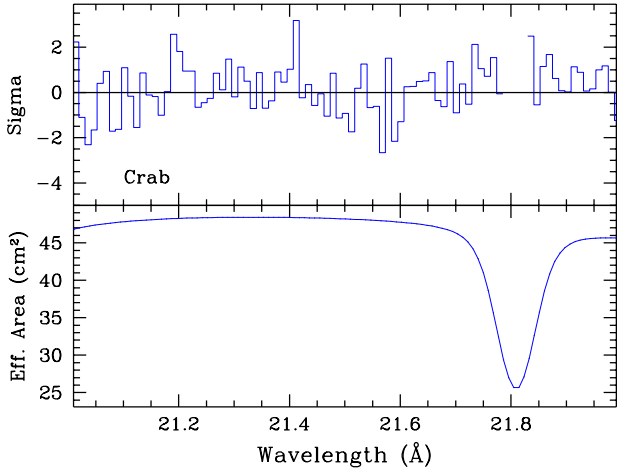,width=8.5cm}
\caption{Residuals left by the best--fit power--law model 
to the [21--22] \AA\, RGS--1 spectrum of the Crab (upper panel).
In the lower panel we show the relative RGS--1 effective area.}
\label{2122-crab} 
\end{center}
\end{figure}

Since the ToO and the archival spectra do not display
significant features in the [20--21] \AA\, range, 
we now concentrate on the [21--22] \AA\, interval, where 
line detections have been claimed by other authors.
In Fig. \ref{2122-spec} we show the [21--22] \AA\, RGS--1 data
of the ToO (left) and of the archival observations (right).
Negative residuals are present
at $\sim 21.6$ \AA\, (label 7, $\sim 2-3 \, \sigma$) and at $\sim 21.8$ \AA\, 
(label 8, $\gtrsim 2 \, \sigma$) both in the ToO and in the archival spectra. 
We reproduced the spectra with a power--law model
plus 2 Gaussians. As before, we fitted the ToO spectrum alone,
then the four archival spectra simultaneously. Finally, we fitted all
the spectra together. In Table \ref{2122-tab} we give
the two Gaussian best--fit parameters.\\
\begin{table}
\begin{center}
\begin{tabular}{cccc}
\multicolumn{4}{c}{RGS--1 [21--22.2] \AA}\\
\hline
Feature & $\lambda$ & FWHM  & k  \\
number  &  (\AA)    & (m\AA) & (cts cm$^{-2}$ s$^{-1}$ \AA$^{-1}$)\\
\hline
\multicolumn{4}{c}{ToO spectrum}\\
\hline
7 & $21.5915\pm0.013$ & $78^{+28}_{-25}$ & $-0.0025\pm0.0007$ \\
8 & $21.801\pm0.0005$ & $7\pm1$ & $-0.018\pm0.005$ \\
\hline
\multicolumn{4}{c}{Archival data}\\
\hline
7 & $21.612\pm0.010$ & $68^{+19}_{-16}$ & $-0.012\pm0.002$ \\
8 & $21.819\pm0.001$ & $12\pm2$ & $-0.342\pm0.068$ \\
\hline
\multicolumn{4}{c}{Total}\\
\hline
7 & $21.603\pm0.008$ & $81^{+17}_{-15}$ & $-0.109\pm0.019$ \\
8 & $21.823\pm0.001$ & $6\pm2$ & $-0.277\pm0.059$ \\
\hline 
\end{tabular}
\caption{Best--fit parameters of the power--law plus two Gaussians
model reproducing the [21--22] \AA\, Mkn 421 spectra taken by the RGS--1.
The errorbars refer to the 90\% confidence intervals.}
\label{2122-tab}
\end{center}
\end{table}
Because  of readout failures in the RGS--2 CCD--4  we cannot
compare the RGS--1 and the RGS--2 data.  To check the reality of the two 
RGS--1 features, we investigated the spectrum of the Crab nebula.
In  Fig. \ref{2122-crab} we show the residuals left by a power--law model
in the RGS--1 Crab spectrum (upper panel) and the effective area
of the instrument  (lower panel).\\
We summarize the results for this energy range as follows:
\begin{itemize}
\item Feature 7 (21.603 \AA):  broad residuals can be observed
in each Mkn 421 spectrum at the observer--frame wavelength
of the OVII K$\alpha$ transition.
The line is not present in the Crab spectrum. We believe it is
caused by an astronomical OVII K$\alpha$ absorbing system. 
\item Feature 8 (21.823 \AA): large residuals can be 
observed at this wavelength (up to $\sim 4\, \sigma$, December 1$^{\rm st}$).
 Unfortunately, the comparison with the Crab data 
is useless because of a large effective area structure.
In previous papers, this line was usually interpreted 
as an OVII K$\alpha$ absorbing system at redshift $z\sim 0.01$.
With the present data, however,  we cannot draw firm conclusions
about its astronomical origin.
\end{itemize}
\subsection{The [22.5--24.5] \AA\, spectra}
\begin{figure}
\begin{center}
\psfig{figure=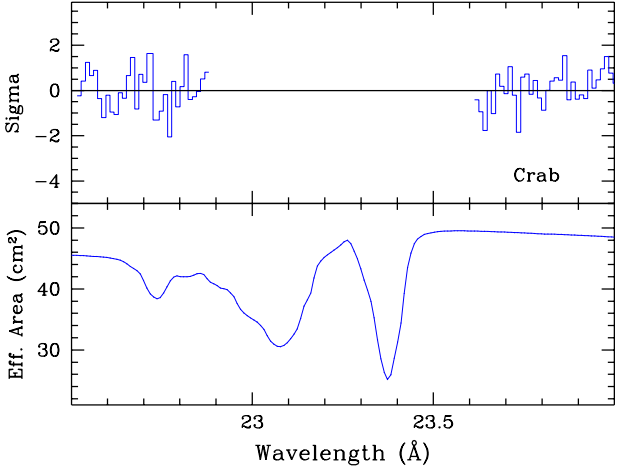,width=8.5cm}
\caption{Residuals left by the best--fit power--law model 
to the [22.5--24] \AA\, RGS--1 spectrum of the Crab (upper panel).
In the lower panel we show the corresponding RGS--1 effective area.
We did not display the residuals in the [22.9--23.6] \AA\, range
which is affected by large instrumental features.}
\label{2224-crab} 
\end{center}
\end{figure}

In Fig. \ref{2224-calib} we showed the [22.5--24.5] \AA\, RGS--1 spectra
of the ToO (left) and of the archival observations (right).\\
This energy range is characterized by the presence of broad instrumental
features caused by oxygen absorption as well as by the interstellar 
absorption around the oxygen K edge (see e.g. de Vries 2003). 
Combining  several Mkn\,421 and PKS\,2155-304
RGS--1 spectra and comparing them with strongly absorbed  Galactic sources,
de Vries et al. (2003) showed the presence of instrumental features
around 23.05 \AA\, and 23.35\, \AA.
The updated calibration files we used to analyze our Mkn 421  data 
account for these structures (see the large features in the RGS--1 
effective areas of Fig. \ref{2224-calib}), 
even if large residuals are still present  at $\sim23.35$ \AA,  
probably caused by uncertainties in calibrating the  instrumental
molecular oxygen absorption (de Vries 2003).\\
We observed two features at a wavelength where the effective areas are smooth,
one of which (label 10 of Fig.\ref{2224-calib}) is the already--discussed 
interstellar neutral oxygen 1s--2p absorption line at $\sim 23.5$ \AA\,
(see Section 4.0.1).
We fitted the [22.5--24.5]\, \AA\, spectra with a power--law + 2 Gaussians
models to reproduce the residuals and we report the results 
in Table \ref{2224-tab}.
\begin{table}[!b!]
\begin{center}
\begin{tabular}{cccc}
\multicolumn{4}{c}{RGS--1 [22.5--24.5] \AA}\\
\hline
Feature & $\lambda$ & FWHM  & k  \\
number  &  (\AA)    & (m\AA) & (cts cm$^{-2}$ s$^{-1}$ \AA$^{-1}$)\\
\hline
\multicolumn{4}{c}{ToO observation}\\
\hline
9 & $22.770\pm0.001$ & $9\pm1$ & $-0.011\pm0.003$ \\
10 & $23.510\pm0.007$ & $64^{+13}_{-12}$ & $-0.004\pm0.001$ \\
\hline
\multicolumn{4}{c}{Archival data}\\
\hline
9 & $22.789^{+0.003}_{-0.002}$ & $15.5\pm5$ & $-0.225\pm0.0665$ \\
10 & $23.510\pm0.013$ & $64^{+26}_{-20}$ & $-0.082\pm0.029$ \\
\hline
\multicolumn{4}{c}{Total}\\
\hline
9 & $22.778\pm0.001$ & $6^{+2}_{-1}$ & $-0.304\pm0.086$ \\
10 & $23.510\pm0.007$ & $71^{+13}_{-11}$ & $-0.133\pm0.020$ \\ 
\hline
\end{tabular}
\caption{Best--fit parameters of the power--law plus three Gaussians
model reproducing the [22--24] \AA\, Mkn 421 spectra taken by the RGS--1.
The error bars refer to the 90\% confidence intervals.}
\label{2224-tab}
\end{center}
\end{table}
As a comparison, we show in Fig. \ref{2224-crab} the residuals left 
by a power--law model of the Crab spectrum. In this case, 
the extremely large count rate strongly suggests calibration uncertainties 
and we therefore avoide reproducing the [22.9--23.6] \AA\, residuals.
We summarize the results as follows:
\begin{itemize}
\item Feature 9 (22.778 \AA): each Mkn 421 spectrum display
residuals ($2-3 \, \sigma$) near a small structure of the RGS--1 
effective areas (at $\sim 22.75$ \AA).
This feature cannot be observed in the Crab spectrum.
In a previous paper, de Vries et al. (2003) tentatively identified 
this line  as an OIV blend produced in the local intergalactic medium.
\item Feature 10 (23.510 \AA): this line was observed 
in all the RGS--1 spectra and can be identified with the 
interstellar neutral oxygen (1s--2p) absorption line (see 
the wavelength calibration Section). 
\end{itemize}
\begin{table*}
\begin{center}
\begin{tabular}{c|c|c|c|c|c}
\hline
Feature & RGS--1           & RGS--2   & Crab  & Literature & Other \\
number & $\lambda$ (\AA)  &          &       &            & sources \\
\hline
2 & $18.680 \pm 0.004$ & $18.697$ (ToO)     & $18.674^{+0.002}_{-0.003}$ & YES (C01--XMM) & PKS 2155-304 (N02--Cha)\\
  &                    & $18.718$ (4/11/02) &                            &  NO (N01--Cha) & PKS 2155-304 (C04--XMM)\\
  &                    & instrumental       &                            & YES (N03--Cha) & \\
\hline
3 & $18.953 \pm 0.003$ & NO                 & instrumental               & YES (C01--XMM) & PKS 2155-304 (N02--Cha)\\
  &                    &                    &                            &  NO (N01--Cha) & 3C 273 (F03--Cha)\\
  &                    &                    &                            & YES (N03--Cha) & NO PKS 2155-304 (C04--XMM)\\
\hline
5 & $19.676 \pm 0.003$ & NO                 & NO                         &  NO            & NO \\
\hline
6 & $19.904 \pm 0.004$ & instrumental       & instrumental               &  NO            & PKS 2155-304 (C04--XMM)\\
  &                    &                    &                            &                & NO PKS 2155-304 (N02--Cha)\\
\hline
7 & $21.603 \pm 0.008$ & no data            & NO                         & YES (C01--XMM) & PKS 2155-304 (N02--Cha)\\
  &                    &                    &                            &  NO (N01--Cha) & PKS 2155-304 (C04--XMM)\\
  &                    &                    &                            & YES (N03--Cha) & 3C 273 (F03--Cha)\\
\hline
8 & $21.823\pm0.001$   & no data            & instrumental               & YES (C01--XMM) & PKS 2155-304$^a$ (XMM)\\
  &     	       &                    &                            & YES$^b$ (N01--Cha) & NO PKS 2155-304 (C04--XMM) \\
  &                    &                    &                            & YES (N03--Cha) & \\		
  &                    &                    &                            &                & \\
\hline
9 & $22.778\pm0.001$   & no data            & NO                         & YES (dV03--XMM)& PKS 2155-304 (dV03--XMM)\\
  &                    &                    & near                       &                & PKS 2155-304 (dV03--Cha)\\
  &       	       &                    & instrumental               &                & PKS 2155-304 (C04--XMM)\\
  &                    &                    & feature                    &                & NO ScoX1(dV03--XMM) \\
  &                    &                    &                            &                & NO 4U 0614+91 (dV03--XMM)\\
\hline       
10 & $23.510\pm0.007$  & no data            &                            & YES (dV03--XMM)& PKS 2155-304 (P01--Cha) \\
   &                   &                    &                            &                & PKS 2155-304 (N02--Cha)\\
   &                   &                    &                            &                & PKS 2155-304 (C04--XMM)\\ 
\hline
\end{tabular}
\caption{Summary of absorption features observed in the RGS spectrum of Mkn421.
$^a$: archival data analyzed by us. 
$^b$: detected in one of the two observations.
C01: Cagnoni (2001); N01: Nicastro et al. (2001); P01: Paerels et al. (2001);
N02: Nicastro et al. (2002); N03: Nicastro et al. (2003);
dV03: de Vries (2003); F03: Fang et al. (2003); C04: Cagnoni et al. (2004).}
\label{tab3}
\end{center}
\end{table*}
\section{Discussion}
In the previous Sections we showed the presence of several absorption
features in the X--ray spectra of Mkn 421
taken with the RGS--1 aboard XMM--Newton.
The X--ray spectra are similar, displaying the same 
absorption lines during all  5 observations over one year.
We discarded some of the lines because they are
very likely instrumental. Among the rest, we found
the well--known interstellar neutral oxygen 1s--2p line 
at $\sim23.5$ \AA\, (see e.g. de Vries et al. 2003).
We observed features at the observer--frame wavelengths of the
expected WHIM lines, i.e. at $\sim 18.6$ \AA\,
(OVII K$\beta$), at $\sim 18.95$\, \AA\, (OVIII K$\alpha$) and  
at $\sim 21.6$ \AA\, (OVII K$\alpha$), which were also reported in some
previous works (e.g. Cagnoni 2001; Nicastro 2003).
We cannot confirm the detection of
the local NeIX absorption feature at $\sim 13.4$ \AA\, 
(Cagnoni 2001; Rasmussen et al. 2003).
We also found two small features, at $\sim 21.8$ \AA\, 
and at $\sim22.78$ \AA, which were identified  in previous papers
as a $z \sim 0.01$ OVII K$\alpha$ line 
(Cagnoni 2001; Nicastro 2003) and as a local  OIV blend 
(de Vries et al. 2003), respectively.\\
However, the astronomical origin of some of these lines is doubtful.
The comparison of the RGS--1 data with the corresponding 
RGS--2 spectra (where available), with a RGS--1 spectrum of the Crab nebula
and with the literature data suggests that some of these lines are probably
instrumental. The absence of an RGS--1 feature in the corresponding
RGS--2 spectrum strongly points towards an instrumental origin of the line. 
Also the literature data are not conclusive:
none of the  reported lines were always detected during the XMM--Newton and
the Chandra observations. We resume these comparisons in Table \ref{tab3}.\\
The suspicions about the astronomical nature of some
of the lines are furtherly strengthened
by their extreme narrowness (see Table \ref{resol}). 
The features at $18.95$ \AA, $\sim 21.8$ \AA\, and at $\sim 22.78$ \AA\, 
are much  smaller (by  a factor $\gtrsim 4$) than the RGS--1 spectral resolution 
($\lambda/\Delta \lambda_{\rm FWHM} < 500$; see XMM--Newton users' handbook).
The feature at $\sim 18.6$ \AA\, is marginally consistent
with the instrumental performances only during the ToO observation.
Even in this case, however, both the corresponding 
RGS--2 line and the feature in the Crab spectrum 
are narrower than the RGS response
($\lambda/\Delta \lambda_{\rm FWHM} > 1000$), supporting 
an instrumental nature.
We therefore conclude that while the features at $\sim 21.6$ \AA,
at $\sim 23.5$ \AA\, can be identified as astronomical lines,
all the others are probably of instrumental origin 
(although some of them have been reported as real astronomical 
lines in previous works by other authors). \\
\begin{table}
\begin{center}
\begin{tabular}{cc|ccc}
\hline
Feature & $\lambda$ & \multicolumn{3}{c}{$\lambda/\Delta \lambda_{\rm FWHM}$} \\
number      & (\AA)   & ToO & Archival & Total \\ 
\hline 
2  & 18.680 & $780^{+190}_{-290}$ & $1230^{+160}_{-240}$ & $570^{+120}_{-150}$ \\ 
3  & 18.953 & $2370\pm300$ & $1310\pm 180$ & $680^{+100}_{-120}$ \\
7  & 21.603 & $280^{+90}_{-100}$ & $320^{+70}_{-90}$ & $270\pm50$ \\
8  & 21.823 & $3110\pm440$ & $1820\pm300$ & $3640\pm1200$ \\ 
9  & 22.778 & $2530\pm280$ & $1470\pm470$ & $3800^{+630}_{-1260}$ \\
10 & 23.510 & $370\pm70$ & $370^{+110}_{-150}$ & $330^{+50}_{-60}$ \\
\hline
\end{tabular}
\caption{$\lambda/\Delta \lambda_{\rm FWHM}$ ratios for the 
possible astronomical features observed in the RGS--1 spectra of Mkn 421.
The instrumental resolving power is $\lambda/\Delta \lambda_{\rm FWHM} < 500$.
}
\label{resol}
\end{center}
\end{table}

The feature located at $23.510\pm0.007$ \AA\, is the well--known interstellar
neutral oxygen (1s--2p) absorption line  which we discussed 
in a previous Section. The feature at $21.603\pm0.008$ \AA\,
is very close to the 
OVII K$\alpha$ transition ($\lambda=21.602$ \AA\, in the observer frame),
the strongest WHIM signature predicted by the simulations
(e.g. Hellsten, Gnedin \& Miralda--Escud\'e 1998).
An OVII absorbing system was already observed 
at zero redshift toward Mkn 421 (Cagnoni 2001; Nicastro 2003)
as well as toward other sources
as PKS 2155-489 (Nicastro et al. 2002; Cagnoni et al. 2004),
3C\,273 (Fang et al. 2002) and H\,1821+643 (Mathur et al. 2003),
while Mckernan et al. (2004) studied the sightlines toward 15 AGNs,
finding evidence of local hot gas in various cases.
This has been attributed either to the WHIM within the local
group of galaxies (e.g. Nicastro et al. 2002, McKernan et al. 2004)
or to radiatively cooling gas inside our Galaxy 
(e.g. Heckmann et al., 2002; McKernan et al. 2004).\\
The average equivalent width of the $21.603$ \AA\, line 
is EW=$9.6^{+3.9}_{-3.1}$ m\AA,  consistent with that found by Cagnoni (2001) 
in her composite spectrum (EW=$12.67^{+1.59}_{-1.54}$ m\AA).
Since the equivalent width of our OVII K$\alpha$ line falls in
 the linear branch of the curve of growth 
calculated by Nicastro et al. (2002), 
it should be produced in an  unsaturated absorption regime.
Using  Fig. 4 of Nicastro et al. (2002), we can therefore calculate 
the column density of OVII toward Mkn 421,
 N$_{\rm OVII} \sim 4\times 10^{15}$ cm$^{-2}$.
A similar result is also obtained using the curves of growth
 calculated by Mathur et al. (2003): the OVII column density 
is in the range $[4-10]\times 10^{15}$ cm$^{-2}$,
depending on the assumed velocity parameter of the gas.\\
Assuming an upper limit to the equivalent width 
of the non--detected OVIII K$\alpha$, we could set an upper limit 
to the temperature of the absorbing gas.
If a line is not saturated, as in our case, the 
equivalent width produced by an ion $X^i$ can be written as
EW$(X^i) \propto A(X) n_{X^i}$ (Nicastro et al. 1999b),
where $A(X)$ is the relative abundance of the element $X$ compared to H
and $n_{X^i}$ is the relative density of the ion $i$ of the element $X$.
We assumed therefore that  the OVIII K$\alpha$ at $\sim 18.97$ \AA\, 
was characterized by the same FWHM of the observed OVII K$\alpha$ 
(FWHM = $0.081^{+0.017}_{-0.015}$ \AA)
and by an amplitude of three times the uncertainty on the continuum 
at the line wavelength. We obtained an average EW=$1.3\pm0.3$ \AA.
Using the OVIII/OVII ratio 
vs temperature $T$ plot  calculated by Nicastro et al. (2002), we obtained
an upper limit for the gas temperature
$T\lesssim1.6\times10^6$ K in the case of a Galactic density
($n_e=1$ cm$^{-3}$). In the case of the extreme extragalactic low-denisity
solution,
$n_e=10^{-6}$ cm$^{-3}$, the gas temperature would be much lower than $10^5$ K, 
while for higher density values, up to $n_e= 5 \times 10^{-5}$ cm$^{-3}$
(Nicastro et al. 2005a), the gas temperature would be compatible with 
values of the order of few times $10^5$ K. Heckman et al. (2002) suggested 
that the zero redshift X--ray absorption lines observed toward PKS 2155-304 
by Nicastro et al. (2002) originated in a radiatively cooling gas inside our Galaxy. 
This would be the case also for Mkn 421 if we assume a low extragalactic density.
The gas would then be too cool to be identified as the searched-for WHIM, 
whose temperature is predicted to be greter than $10^5$ K 
(e.g Hellsten, Gnedin \& Miralda--Escud\'e 1998). However, if we assume
a larger density value, then our data are still compatible with a WHIM
with a temperature of the order of a few times $10^5$ K.
\section{Conclusions}
We presented the analysis of the data of the high resolution 
RGS spectrometers aboard XMM--Newton during a ToO observation of Mkn 421
performed in November 2003. The pointing was triggered since the source was
in a high state of activity in the X--ray band. 
We compared these data with 3 archival RGS observations of the same source
performed in November and December 2002 and with one performed in June 2003.
We summarize the main results:

\begin{itemize}

\item The integrated [7--36]\, \AA\, RGS--1 and RGS--2 spectra 
are well fitted by a broken power--law model, with a hard spectral index 
($\alpha_1\sim0.6-0.9$) below 0.7--0.8 keV softening toward
higher energies ($\alpha_2\sim 1.0-1.3$). The RGS--2 
spectra show systematically softer slopes and lower fluxes.

\item We compared the RGS spectra to the corresponding EPIC--PN spectra 
in the common well--calibrated energy range: [0.6--1.77] keV.
The EPIC--PN spectra are significantly softer, always showing 
spectral indices $\alpha>1$, with fluxes larger by 5--10\%.

\item We focused on small sections, 2--3 \AA\, wide, of the RGS--1 spectra,
looking for absorption features. We found several lines
which are common to all the observed spectra. However, we found evidence
that several of them  are instrumental: their proximity to 
effective area features, their absence in the  corresponding  
RGS--2 spectra or their extreme narrowness 
(much smaller than the instrumental resolution).
We found that only two features are very likely of astronomical origin:
one at $\sim 23.5$ \AA\, and one at $\sim 21.6$ \AA.

\item We identified the feature at $23.5$ \AA\, as  interstellar
neutral oxygen absorption, as proposed e.g. by de Vries et al. (2003).
We found a best--fit wavelength of $23.510\pm0.007$ \AA, 
slightly larger than the theoretical position proposed by
Mc Laughlin \& Kirby (1998), but fully consistent with the experimental
results of several authors.

\item The feature at $\sim 21.6$ \AA\, corresponds to a zero--redshift
OVII K$\alpha$ transition. Assuming a 3$\sigma$ upper limit on 
the non--detected corresponding OVIII K$\alpha$, 
we calculated an upper limit for the gas temperature. 
We found that an extragalactic gas with a very low density
would be much colder than the predicted WHIM temperature range.
In this case  this line would have a Galactic origin, produced by 
a radiatively cooling gas, as proposed by Heckman et al. (2002).
For an extragalactic gas with higher density, WHIM with a temperature 
of a few times $10^5$ K would still be compatible with the data.

\end{itemize}

We conclude from our analysis that, with the current knowledge of the
XMM-Newton grating performances and the sensitivity provided by our spectra,
we could not find firm evidence of the Warm/Hot Intergalactic Medium toward Mkn 421.

\begin{acknowledgements}
We thank the referee for useful comments that helped us to improve
the paper. This research was finacially supported by the Italian Ministry 
for University and Research.  
\end{acknowledgements}

\begin{noteaddname}
After this paper was accepted, the Nicastro et al. (2005b) paper
appeared in the literature. Nicastro et al. were able to observe
Mkn\,421 in much higher states with the grating on board the Chandra
satellite. Thanks to the higher signal to noise ratio they were able to
identify many more astronomical lines that allow them to derive for the
first time a population of baryons in two intervening WHIM systems
at $z=0.011$ and  $z=0.027$ and to study in detail the zero--redshift
system (Williams et al. ApJ submitted).
\end{noteaddname}

\appendix{

\section{RGS spectral analysis and comparison with EPIC-PN}

First, we analyzed the RGS--1 and RGS--2 spectra 
in the full [0.34--1.77] keV energy range  ([7--36] \AA), 
where the calibration uncertainties are smaller than 10\%.
We rebinned the RGS data  to have at least 2000 counts per bin
and added a systematic error of $3\%$.
We fitted the data with an absorbed power--law 
and a broken power--law model, 
fixing the absorption parameter to the Galactic value
(N$_{\rm H} =1.61\pm 0.1 \times 10^{20}$ cm$^{-2}$; Lockman \& Savage 1995).
In Table \ref{rgs-bband-spec} we report the best--fit parameters 
of each observation. The RGS--2 spectra display systematically steeper 
slopes and lower fluxes than the RGS--1, as reported in the calibration 
paper on the radio--loud narrow--line Seyfert 1 galaxy  PKS 0558-508
(Kirsch 2003).

\begin{figure*}
\begin{center}
\hbox to \textwidth
{
\vbox {\psfig{figure=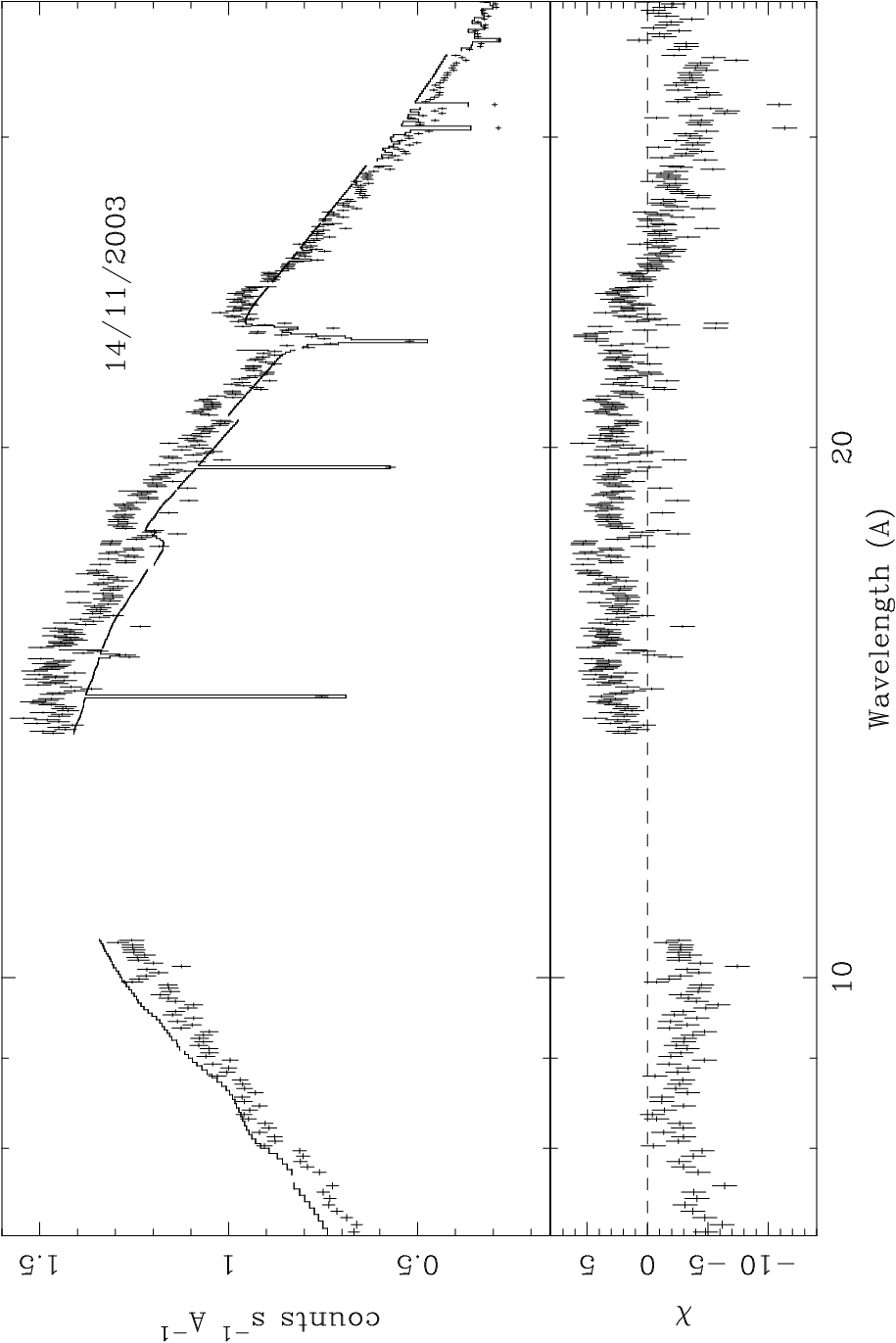,width=7.5cm,angle=-90}}
\hskip 0.3cm
\vbox{\psfig{figure=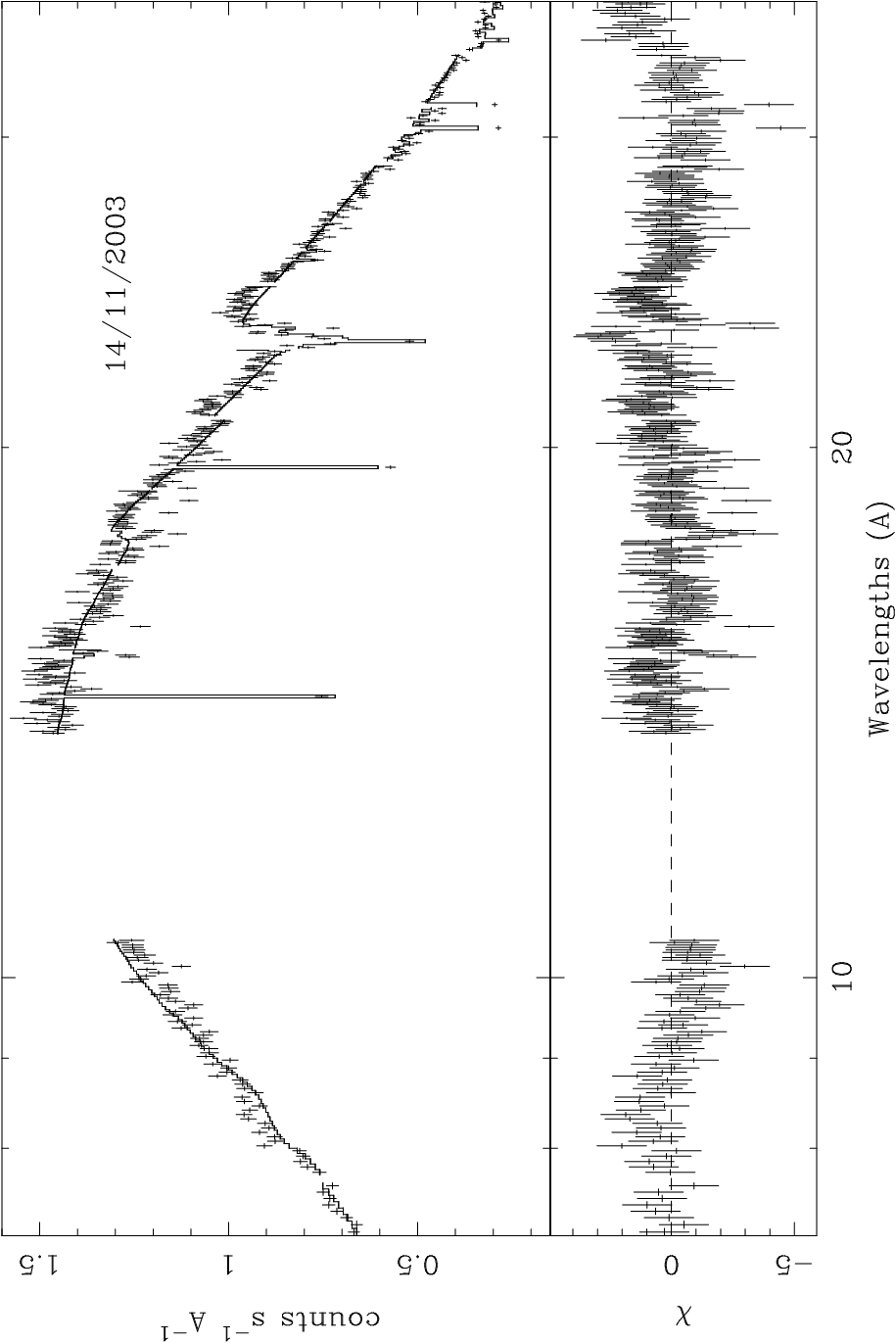,width=7.5cm,angle=-90}}
\hfill
}
\caption{The RGS--1 spectrum of the ToO observation of November 14, 2003.
An absorbed power--law model is not able to  reproduce the data well
({\it left panel}), while a broken power--law model fits 
the data better ({\it right panel}).
We fixed the absorption parameter to the Galactic value.
The spectra were rebinned in order to have at least 2000 counts per bin.
The other RGS spectra display similar behaviors.}
\label{rgs-total-spec}
\end{center}
\end{figure*}

\begin{table*}
\begin{center}
\hspace{-0.5cm}
\begin{tabular}{ccccccc}
\hline
 Inst. & $\alpha_1$ & E$_b$ & $\alpha_2$ &  F$_{1 keV}$ & F$_{0.5-1.5keV }$ & $\chi^2_r/d.o.f.$ \\
     &            & (keV) &            & ($\nu$Jy) & ($\times 10^{-10}$)$^a$  & \\ 
\hline
\multicolumn{7}{c}{Exp. 0136540301}\\
\hline
 RGS--1 & $0.99$ &  & & 101.9 & 2.71 & 3.26/127\\
 RGS--1 & $0.755^{+0.055}_{-0.035}$ & $0.74^{+0.1}_{-0.03}$ & $1.27^{+0.11}_{-0.04} $ & 104.9 & 2.72 & 0.91/125 \\
 RGS--2 & $1.05$ & & & 92.4 & 2.48 & 4.24/125\\
 RGS--2 & $0.81\pm0.03$ & $0.80^{+0.02}_{-0.03}$ & $1.46^{+0.04}_{-0.06}$ & 97.3 & 2.51 & 0.90/125 \\  
\hline
\multicolumn{7}{c}{Exp. 0136540401}\\
\hline
 RGS--1 & $0.93$ & & & 120.1 & 3.16 & 3.89/137\\
 RGS--1 & $0.63^{+0.03}_{-0.06}$ & $0.69^{+0.02}_{-0.04}$ & $1.20^{+0.03}_{-0.04}$ & 123.2 & 3.19 & 0.94/135\\
 RGS--2 & $0.94$ & & & 115.9 & 3.05 & 2.52/144\\
 RGS--2 & $0.77^{+0.04}_{-0.02}$ & $0.79^{+0.03}_{-0.05}$ & $1.18\pm0.04$ & 121.0 & 3.07 & 1.21/142\\
\hline
\multicolumn{7}{c}{Exp. 0136541001}\\
\hline
RGS--1 & $0.90$ & & & 72.8 & 1.91 & 2.83/227\\
RGS--1 & $0.64\pm0.02$ & $0.67^{+0.03}_{-0.02}$ & $1.11\pm0.02$ & 75.0 & 1.92 & 0.92/225\\
RGS--2 & $0.92$ & & & 68.7 & 1.81 & 2.5/240\\ 
RGS--2 & $0.74^{+0.03}_{-0.02}$ & $0.76^{+0.03}_{-0.04}$ & $1.17^{+0.02}_{-0.04}$ & 71.6 & 1.82 & 0.97/238\\
\hline
\multicolumn{7}{c}{Exp. 0158970101}\\
\hline 
RGS--1 & 1.06 & & & 74.0 & 1.98 & 3.5/138\\
RGS--1 & $0.80\pm0.04$ & $0.71^{+0.02}_{-0.04}$ & $1.33^{+0.03}_{-0.04}$ & 75.3 & 2.00 & 1.16/136\\
RGS--2 & 1.10 & & & 68.5 & 1.85 & 2.3/139\\
RGS--2 & $0.92^{+0.03}_{-0.04}$ & $0.76\pm0.05$ & $1.35\pm0.04$ & 71.2 & 1.86 & 0.83/137\\
\hline
\multicolumn{7}{c}{Exp. 0150498701}\\
\hline
RGS--1 & $0.80$ & & & 186.5 & 4.83 & 3.17/454\\
RGS--1 & $0.56\pm0.02$ & $0.71\pm0.02$ & $1.015\pm0.015$ & 192.4 & 4.87 & 1.13/452\\
RGS--2 & 0.82 & & & 174.3 & 4.52 & 2.42/490\\
RGS--2 & $0.64\pm0.02$ & $0.73\pm0.02$ & $1.02^{+0.01}_{-0.02}$ & 179.8 & 4.56 & 1.08/488\\
\hline
\multicolumn{7}{c}{Parabolic model}\\
\hline
   & a &  & b &   &  &  \\
\hline
\multicolumn{7}{c}{Exp. 0136540301}\\
\hline
RGS--1 & $1.21^{+0.02}_{-0.03}$ & & $0.77\pm0.08$ & 110.9 & 2.74 & 1.02/126\\
RGS--2 & $1.33^{+0.02}_{-0.03}$ & & $0.96^{+0.06}_{-0.1}$ & 101.2  & 2.52 & 1.1/126\\
\hline
\multicolumn{7}{c}{Exp. 0136540401}\\
\hline
RGS--1 & $ 1.17^{+0.02}_{-0.03}$ & & $0.86^{+0.05}_{-0.09}$ & 131.5 & 3.21 & 1.11/136\\
RGS--2 & $1.10^{+0.02}_{-0.03}$ & & $0.58^{+0.05}_{-0.09}$ & 125.4  & 3.08 & 1.36/143\\
\hline
\multicolumn{7}{c}{Exp. 0136541001}\\
\hline
RGS--1 & $1.09^{+0.01}_{-0.04}$ & & $0.67^{+0.04}_{-0.07}$ & 79.3 & 1.93 & 1.07/226\\
RGS--2 & $1.10^{+0.01}_{-0.02}$ & & $0.60^{+0.05}_{-0.06}$ & 74.5 & 1.83 & 1.13/239\\
\hline
\multicolumn{7}{c}{Exp. 0158970101}\\
\hline
RGS--1 & $1.28^{+0.02}_{-0.03}$ & & $0.77^{+0.06}_{-0.08}$ & 80.6 & 2.01 & 1.33/137\\
RGS--2 & $1.29\pm0.02$ & & $0.65^{+0.06}_{-0.09}$ & 74.3 & 1.87 & 0.91/138\\
\hline
\multicolumn{7}{c}{Exp. 0150498701}\\
\hline
RGS--1 & $0.96\pm0.01$ & & $0.65^{+0.05}_{-0.03}$ & 204.0 & 4.91 & 1.25/453\\ 
RGS--2 & $0.97^{+0.01}_{-0.02}$ & & $0.53^{+0.03}_{-0.04}$ & 188.6 & 4.57 & 1.25/489\\
\hline
\end{tabular}
\caption{Best--fit parameters of the RGS spectra reproduced
with power--law, broken power--law and parabolic models.
We fixed the absorption parameter to the Galactic value.
The error bars refer to the 90\% confidence interval.
$^a$: erg cm$^{-2}$ s$^{-1}$.
}
\label{rgs-bband-spec}
\end{center}
\end{table*}

\begin{figure}
\begin{center}
\psfig{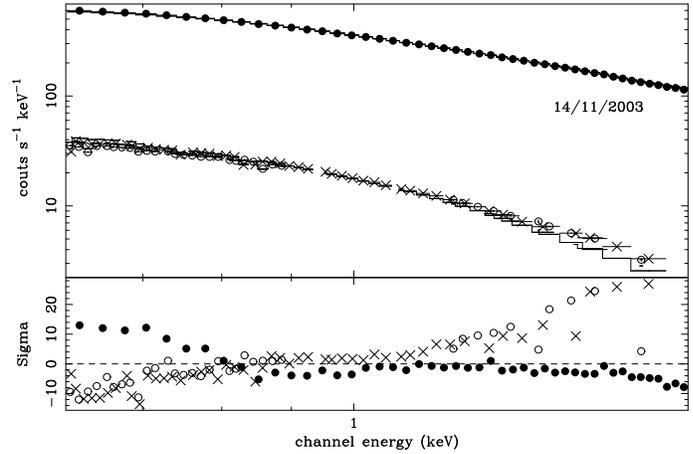}
\caption{EPIC--PN (filled circles), RGS--1 (empty circles)
and RGS--2 (crosses) spectra of Mkn 421
during the ToO observation of November 14, 2003, fitted together
with a broken power--law model (with Galactic absorption value).
For plotting purposes, we rebin the  the RGS data to have 
at least 15000 counts in each bin.
The  RGS spectra are significantly harder than the EPIC--PN spectrum.
For clarity, we omitted the residual error bars.}
\label{pn+rgs-spec}
\end{center}
\end{figure}

The power--law model  cannot reproduce the spectra:
the $\chi^2_r$ are always larger than 2 (e.g. Fig. \ref{rgs-total-spec}).
The broken  power--law model significantly improves the quality 
of the fit: the spectra of Mkn 421
are hard up to $\sim 0.7-0.8$ keV and soften toward higher energies.
We also tried to reproduce the steepening with the logarithmic parabolic
model proposed by Massaro et al. (2003a,b), which we already applied 
to the corresponding EPIC--PN spectra (Ravasio et al. 2004), 
as well as to the {\it Beppo}SAX spectra of other 
similar sources, such as 1ES\,1959+650 (Tagliaferri et al. 2003).
This curved model reproduces the data  well, even if,
in all cases, the broken power--law gives better results.

To check  the quality and the good calibration of these data,
we compared them to the corresponding EPIC--PN spectra
in the common well--calibrated energy range ([0.6--1.77] keV).
The RGS and the EPIC--PN data are well fitted by
a broken power--law model softening toward higher energies.
The EPIC--PN spectra are significantly softer, always with slopes $\alpha > 1$,
while the RGS spectra are hard up to $\sim 0.8$ keV.
The greater softness of the EPIC--PN was shown also
by Kirsch (2003), using an XMM--Newton observation of PKS 0558-508.
This is very clear in Fig. \ref{pn+rgs-spec}, where we plot 
the RGS and the EPIC--PN spectra of the ToO observation.
Furthermore, the EPIC--PN  fluxes are larger by $\sim 5-10\%$.
As an example, we report in Table \ref{rgs-pn-tab}  the best--fit parameters
of the power--law and  broken power--law models for the November 2003 ToO observation.
These calibration differences, however, are not important for our search for WHIM
features. We focused on such small sections of the RGS spectra (2--3 \AA\, wide)
that the continuum uncertainties can be neglected. 

\begin{table*}
\begin{center}
\begin{tabular}{ccccccc}
\hline
 Inst. & $\alpha_1$ & E$_b$ & $\alpha_2$ &  F$_{1 keV}$ & F$_{0.6-1.5 keV}$ & $\chi^2_r/d.o.f.$ \\
      &           & (keV) &    & ($\mu$Jy) & ($\times 10^{-10}$)$^a$  & \\ 
\hline
\multicolumn{7}{c}{November 14, 2003}\\
\hline
RGS--1 & $0.97\pm0.01$ & & & 188.3 & 4.17 & 1.13/246 \\
RGS--1 & $0.76\pm0.07$ & $0.80^{+0.05}_{-0.03}$ & $1.03^{+0.03}_{-0.02}$ & 192.3 &  4.18 & 0.91/244\\
RGS--2 & $0.97\pm0.01$ & & & 176.4 & 3.91 & 1.11/330\\
RGS--2 & $0.75^{+0.09}_{-0.12}$ & $0.78\pm0.04$ & $1.02\pm0.02$ & 180.0 & 3.90 & 1.0/328\\
PN & $1.34\pm0.01$ & & & 202.3 & 4.35 & 6.6/45\\
PN & $1.23^{+0.02}_{-0.03}$ & $1.02\pm0.06$ & $1.43\pm0.02$ & 207.2 & 4.35  & 0.81/43 \\  
\hline
\end{tabular}
\caption{Best--fit parameters of the ToO RGS and EPIC--PN spectra
fitted with power--law and broken power--law models, 
in the common energy range ([0.6--1.77] keV).
The absorption parameter was fixed to the Galactic value.
Systematic errors of 3\% and of 0.5 \% were added to the 
RGS and to the  EPIC--PN data, respectively.
The error bars refer to the 90\% confidence interval for one parameter.
 $^a$: erg cm$^{-2}$ s$^{-1}$.}
\label{rgs-pn-tab}
\end{center}
\end{table*}

}
\end{document}